# Feedback Lock-in: A versatile multi-terminal measurement system for electrical transport devices

Arthur W. Barnard[a,c,d*], Evgeny Mikheev[a,b], Joe Finney[a], Han S. Hiller[c], and David Goldhaber-Gordon[a,b*]

[a] Department of Physics, Stanford University, 94305 Stanford, California, USA
[b] SLAC National Accelerator Laboratory, 94025 Menlo Park, California, USA
[c] Department of Physics, University of Washington, 98195 Seattle, Washington, USA
[d] Department of Materials Science and Engineering, University of Washington, 98195 Seattle, Washington, USA

*awb1@uw.edu, *golhaber-gordon@stanford.edu

We present the design and implementation of a measurement system that enables parallel drive and detection of small currents and voltages at numerous electrical contacts to a multi-terminal electrical device. This system, which we term a feedback lock-in, combines digital control-loop feedback with software-defined lock-in measurements to dynamically source currents and measure small, pre-amplified potentials. The effective input impedance of each current/voltage probe can be set via software, permitting any given contact to behave as an open-circuit voltage lead or as a virtually grounded current source/sink. This enables programmatic switching of measurement configurations and permits measurement of currents at multiple drain contacts without the use of current preamplifiers. Our 32-channel implementation relies on commercially available digital input/output boards, home-built voltage preamplifiers, and custom open-source software. With our feedback lock-in, we demonstrate differential measurement sensitivity comparable to a widely used commercially available lock-in amplifier and perform efficient multi-terminal electrical transport measurements on twisted bilayer graphene and SrTiO₃ quantum point contacts. The feedback lock-in also enables a new style of current-biased measurement which we demonstrate on a ballistic graphene device.

## Introduction

Electrical transport measurements play a central role in understanding behavior of electrons in solids. Such measurements involve sourcing currents and/or voltages in particular locations on a "device"—a material or set of materials structured in some well-defined geometry—and probing how the resultant voltages and/or currents measured at the same or different locations within the device vary with external control parameters (e.g. temperature, magnetic field, electric field). When the material is patterned on small length scales—the so-called mesoscopic regime[1]—electrical transport measurements can reflect quantum mechanical wave behavior, ballistic motion, Coulomb interactions, and topological properties that are hidden on familiar macroscopic scales. In this regime, physical phenomena can be highly sensitive to the magnitude of applied currents and voltages, so to probe such devices it is critical to keep these excitations small, which in turn creates a need to measure small signals. To this end, lock-in amplifiers are commonly used to measure small signals amid comparatively large background noise by measuring a narrow-bandwidth response at the frequency of a sinusoidal drive signal.

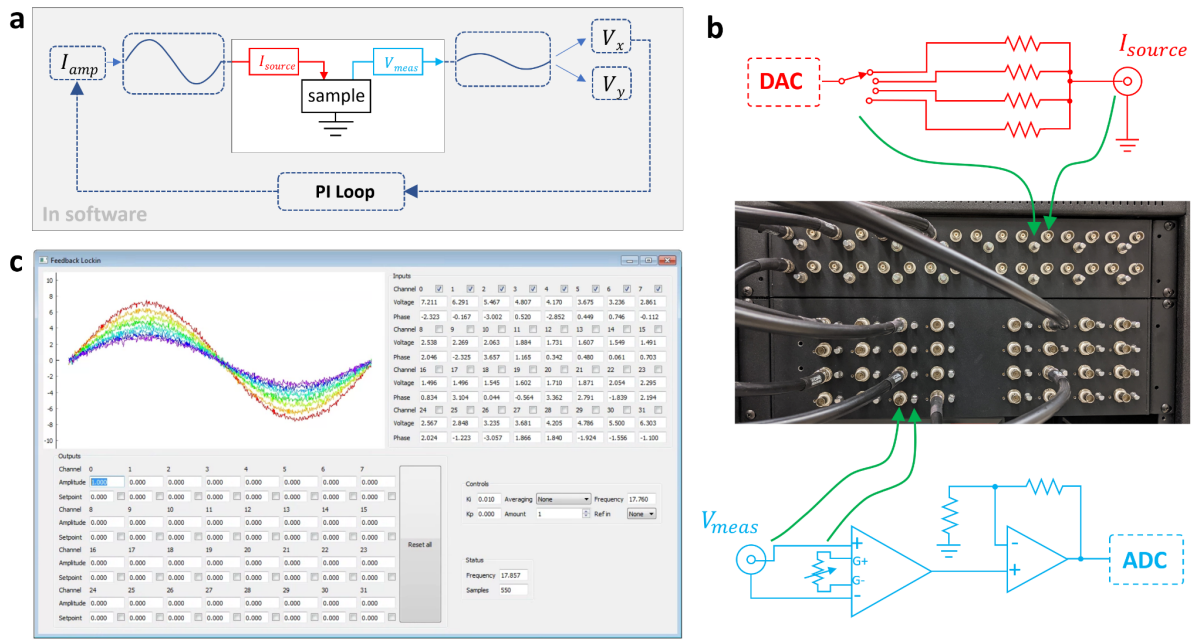

**Figure 1** *Feedback lock-in design.* **(a)** *Schematic operation of feedback lock-in for a single channel on a 2 (or 3) terminal device. Computer generates a sinusoidal drive signal that passes through a DAC and a bias resistor to provide current $I_{source}$ to the sample. Voltage $V_{meas}$ is collected, amplified, and read-in to software via an ADC. The result is multiplied by a sine and cosine wave and integrated to produce $V_x$ and $V_y$. An optional feedback loop can adjust the sourced current to maintain a $V_{meas}$ setpoint.* **(b)** *Hardware implementation. Current source is diagrammed in red and voltage measurement is diagrammed in teal. The 32 outputs each have 4 selectable bias resistors, and the 32 inputs have 3-4 selectable gain settings.* **(c)** *Graphical user interface.*

Lock-in techniques are effective at filtering out noise, however there is an inverse relationship between the amount of filtering and the averaging time needed to perform a single measurement[2]. Thus, transport measurements are inherently speed-limited, and it often takes weeks to months to study a single device, especially if its behavior as a function of multiple parameters is of interest. Given the considerable expense in time and resources to perform such measurements, it is valuable to parallelize data acquisition. Modern condensed matter physics increasingly relies on multi-terminal devices to probe the physics of new materials[3–6], so the ability to efficiently probe 10 or more channels simultaneously can dramatically accelerate the discovery process and avoid requiring experimentalists to pre-select which subset of data to take.

Multi-channel[7–9], low-cost, and computer-based[10–13] lock-in amplifiers have been developed for a variety of applications. Recent developments[14–20], based on commercial FPGAs, compare favorably with common commercial lock-in amplifiers. However, to date such approaches have not made significant impact on studies of mesoscopic devices, in part because a multi-terminal system hasn't been demonstrated in a configuration tailored to quantum transport devices.

Here, we demonstrate such an instrument, which can be built from discrete components and commercially available data acquisition boards, at a total price comparable to that of a single commercial lock-in amplifier. Besides replacing the function of many lock-in amplifiers inexpensively and compactly, this system's capability to programmatically switch the effective impedance of individual measurement leads enables versatile control over the measurement configuration, including accessing configurations that would otherwise be practically impossible to achieve.

**Feedback Lock-in design**

Our measurement system, which we term a feedback lock-in, combines commercial digital acquisition boards (from National Instruments, henceforth NI, in our present design) with home-built voltage amplifier boards and custom open-source software running on a consumer-grade desktop PC. It operates by sourcing sinusoidal currents and measuring the voltage response for each period of the sine wave. The in-phase ($V_x$) and out-of-phase ($V_y$) components are computed and can be averaged or fed back on to adaptively change the source current amplitude. When no feedback is used, the source impedance is naturally high, however when feedback is on, a source can act as a low-impedance virtual ground.

The basic configuration for a single channel measurement is outlined in Fig. 1a,b. A software-generated sinusoidal voltage signal passes through a high-impedance bias resistor to an electrical lead of the device. The voltage response of the same (or different) lead is then read in through a variable gain preamplifier whose output is digitized by a commercial analog-to-digital converter (ADC) channel. The recorded voltages are then multiplied by in-phase and out-of-phase reference sine waves and the results numerically integrated to produce the measured $V_x$ and $V_y$ components. The measured components can then be time-averaged using computer-defined window functions (e.g. exponential or rectangular) to improve the signal-to-noise ratio (SNR). The averaged signal can then be passed to a software-defined digital proportional-integral (PI) loop to enable adaptive control of the source current. In the 2-terminal configuration depicted in Fig. 1a, the feedback loop can be used to effectively enable a voltage-biased measurement (wherein the source current serves as the dependent variable probed). The single channel behavior can be easily generalized, so that one or more currents are sourced, and a large number of voltages are measured simultaneously.

Because much of the feedback lock-in operation is carried out in software, there is flexibility to add functionality that is uncommon in stand-alone lock-ins. For example, the sliding window average is particularly effective for transport measurements, since it allows the user to maximize the number of sinusoidal cycles averaged for each data point without a risk of unintended bleedthrough between data points in exponential averaging (low pass filtering).

**Software interface**

Our open-source homebuilt software interface is written in Python, using the PyVISA driver platform to communicate with NI boards. It includes a simple TCP server, allowing external programs (including those not based on Python) to control and read out the instrument status. An example of the graphical user interface (using PySide2) is shown in Fig 1c. Based on conditions set in a configuration file, the software automatically populates the UI with the appropriate number of input/output channels, enabling designs relying on different NI boards.

In the upper right of Fig. 1c, the amplitude and phase of each of 32 channels is displayed. For each channel, a checkbox toggles whether a real-time oscilloscope trace is plotted in the upper left. In the lower left, 32 outputs' amplitude can be directly set or controlled by a feedback loop setpoint when the associated checkbox is set. In the lower right are global control parameters, including the PI loop constants ($K_i$ and $K_p$), type of averaging, lock-in frequency, and an option to set a reference input signal for the feedback loops. Finally, a "Reset all" button in the lower middle gives the user a quick way of aborting a measurement to avoid damage to a device.

**Performance of feedback lock-in**

To performance-test the feedback lock-in, we measure a surface mount resistor installed on a ceramic chip carrier inside a table-top vacuum cryostat to ensure that typical noise characteristics are present. We first measure the resistance using a calibrated commercial source-measure unit (model Keithley 2450) in 4-wire sense mode. Then, we measure the same resistor using the feedback lock-in where we use two independent inputs for the high and low voltage terminals in a 4-terminal configuration. For the results shown in Fig. 2a, we use a drive frequency of 17.76 Hz and an exponential averaging with a time constant of 300 ms, and sample each configuration once per second for 100 samples. The error bars in Fig. 2a correspond to the standard deviation among the single one-second measurements, and thus represent the measurement uncertainty assuming only one one-second sample per experimental configuration. We use source currents (of order 1 nA) which are at or below typical currents used in transport measurements, and show improved SNR with increasing source current, as expected.

To benchmark the performance of the feedback lock-in, we compare it to a similar measurement performed on a Stanford Research Systems SR830 lock-in amplifier using an SR560 voltage preamplifier. In Fig. 2b, we plot the fractional uncertainty as a function of the measured resistance, using 3 different feedback lock-in configurations and the SR830. The first feedback lock-in configuration (blue circle) is that described above related to Fig. 2a. The second feedback lock-in configuration (red x) improves the SNR by using a combiner and a single input and thus a single differential voltage amplifier. The third configuration also uses a single differential amplifier but in place of an exponential average, uses a 1 second sliding window. We choose this window size since we sample data every 1 s.

In measuring with the SR830, we source the current from the feedback lock-in and deliver a reference signal from the feedback lock-in to the SR830 to eliminate variability in output source noise between instruments. We use an SR560 with a newly installed amplifier chip to ensure that there is no additional technical noise introduced by a degraded amplifier. On the SR560, we employ a differential measurement using a 1000x gain setting, a 0.1Hz-10kHz band pass filter, and a low-noise dynamic reserve setting. On the SR830 we maximize the sensitivity (10 mV) and use a 6 dB/dec low pass filter with a 300 ms time constant. Higher order filters are available on the SRS830 (and could be implemented into the feedback lock-in software) though they require shorter time constants to achieve comparable settling times.

The differential measurement condition (red x) of the feedback lock-in is the most direct comparison to the SR830 + SR560. Both instruments are using low pass filters with similar time constants, and on this metric, the feedback lock-in has comparable performance with approximately 20% higher uncertainty. However, the feedback lock-in's SNR is improved by utilizing a sliding window average. As can be seen, in this condition (orange diamond), the feedback lock-in and SR830+SR560 have nearly equivalent performance, with the feedback lock-in potentially out-performing the SRS830+SR560 at low source currents.

As shown, the feedback lock-in's performance is comparable to that of the SR830 for a single channel in realistic transport measurement conditions. It should be noted that the SR830 has high frequency capabilities (102 kHz) that we have not attempted to implement in the feedback lock-in because transport measurements on semiconductors are almost exclusively performed in the 1-100 Hz range to reduce the impact of finite line impedances. Our implementation multiplexes a 1 MHz sampling clock and consequently has a digital sample rate of 33kHz per channel—meaning that it is plausible to trade off a reduced number of usable channels to access frequencies comparable to those reached by an SR830. Other RF-based methods speed up measurements but require specialized expertise and costly external hardware[21,22].

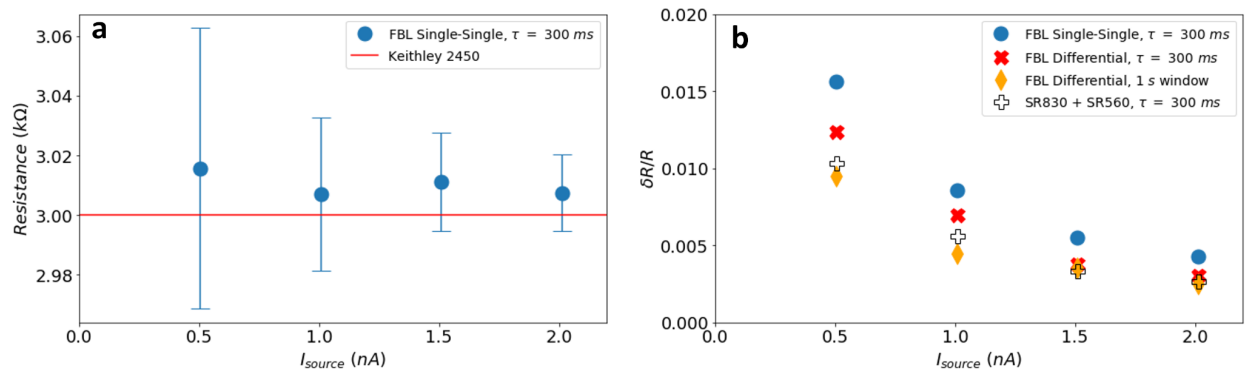

**Figure 2.** *Performance of feedback lock-in (FBL) in realistic conditions.* **(a)** *measurement of 4-terminal resistance of 3 kOhm resistor inside a tabletop cryostat, measured with frequency 17.76 Hz. A reference line for the resistance determined by a calibrated Keithley 2450. The error bars denote the standard deviation $\delta R$ of 100 successive samples.* **(b)** *Measurement modalities compared with a commercial lock-in amplifier. The circle data are obtained by measuring terminals on distinct amplifiers and subtracting the measured results. The x data use a combiner and measure the difference between two terminals using a single differential amplifier. The diamond data use the same measurement conditions as the x data but replace the exponential time average with a windowed time-average. The + denotes the performance of an SR830 lock-in amplifier (6dB/dec filter setting) and SR560 voltage preamplifier. A single channel of the FBL has comparable noise performance to that of the combined SR830 and SR560.*

**Efficiently probing multi-terminal devices**

As discussed above, if we use the feedback lock-in as a conventional lock-in, its per-channel measurement performance is comparable to that of the SR830+SR560. However, the feedback lock-in has not just one but 32 channels. Here we show this parallelism accelerate measurements in two model multi-terminal devices.

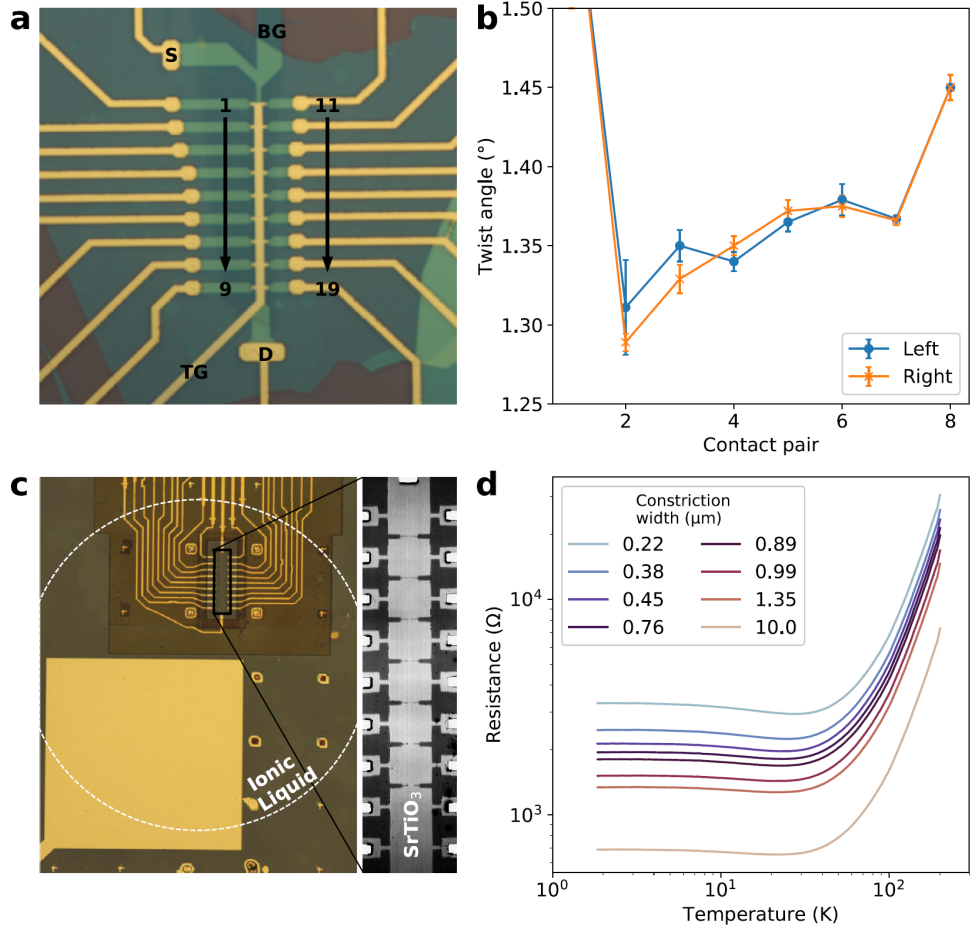

*Figure 3* High throughput measurements on multiterminal devices. *(a,b)* A 22-terminal twisted bilayer graphene heterostructure device[23] has all contact pairs measured simultaneously, providing information on the local twist angle. *(c,d)* A 22-terminal SrTiO3 device[24] with fabricated nano constrictions probes temperature-dependent resistance of 8 constrictions during a single cool-down.

In twisted bilayer graphene with ~1.1deg interlayer twist, a moiré superlattice generates a nearly flat subband in which electronic correlations can have dramatic effects[25,26]. In our and others' work to date, though a specific twist angle between graphene layers is targeted, actual twist angle commonly varies spatially. In the 22-terminal device shown in Fig. 3a[23], as a function of gate voltage we simultaneously measure longitudinal resistance at every pair of neighboring contacts. Noting the carrier density associated with subband filling yields the twist angle in each location, revealing the spatial inhomogeneity in twist angle (Fig. 3b). Also sweeping magnetic field while still simultaneously acquiring data from each contact pair provides important insights into how twist angle impacts the observed physics[23].

Another example of the capability to simultaneously measure many voltage terminals is on nanopatterned 2D electron gases (2DEGs) in $SrTiO_3$. The device shown in Fig. 3c is a long Hall bar where each segment between voltage probes contains a mesoscale constriction of variable width (0.22-1.35 microns), patterned by e-beam lithography and lift-off of sputtered $SiO_2$. A 2DEG is accumulated on the exposed $SrTiO_3$ surface in the Hall bar channel by covering both the device and the adjacent large side gate with an ionic liquid and applying a gate voltage to the side gate.

The lock-in output is used to source an AC current between the top and bottom ends of the Hall bar. The lock-in input channels monitor all 18 voltage probes, measuring 16 voltage differences between adjacent probes (8 voltages on each side of the Hall bar). The measured longitudinal resistance of each segment is shown in Fig. 3d as a function of temperature, with all data taken in a single gradual cooldown. Such measurements allowed us to study the scaling of charge transport as a function of patterned constriction geometry. We use similar multi-terminal Hall bar devices[24] to rapidly assess inhomogeneity in Hall carrier density and mobility, guiding quick iterative adjustments to device fabrication.

In both devices, it would have required a minimum of 16 commercial lock-ins operating in parallel, which is atypical of a transport lab, and prohibitively expensive to acquire for every measurement set-up.

**Novel transport device topologies enabled by feedback lock-in**
The use-case described above shows how the feedback lock-in can be applied directly to existing transport devices to accelerate the measurement process. However, the feedback capability also enables a new type of measurement based on idealized current sources and sinks, which is naturally suited to new device geometries.

The device shown in Fig. 4 is designed to probe the angular flow profile of electrons through a constriction in high-quality hBN-encapsulated monolayer graphene by using 7 drain contacts along the perimeter of a semi-circular device-region. If these drains have minimal contact resistance, the angular current distribution through the constriction can be determined by measuring the fractional current that passes through each drain. If the line/contact resistance is not ideal, then voltage differences among the drain contacts can build-up and obscure the results, even if low-impedance current preamplifiers are used.

To enable idealized current measurements, we designed our device to have two leads connecting to each ohmic contact on the device. Thus, we are able to monitor the voltage of a given contact, while sourcing the current to the contact via a separate lead. Using these voltage probes, the feedback lock-in can then be used to feed back on the measured voltage of all seven contacts to keep them all at the same potential. In our particular measurement, we use one of the drains (contact 4) as a ground reference, and the feedback lock-in maintains a zero voltage difference between each other drain and contact 4. The (nonzero) voltage on the source is also referenced to that of contact 4. Conservation of current allows us to determine the current through contact 4 by summing the current through the other leads.

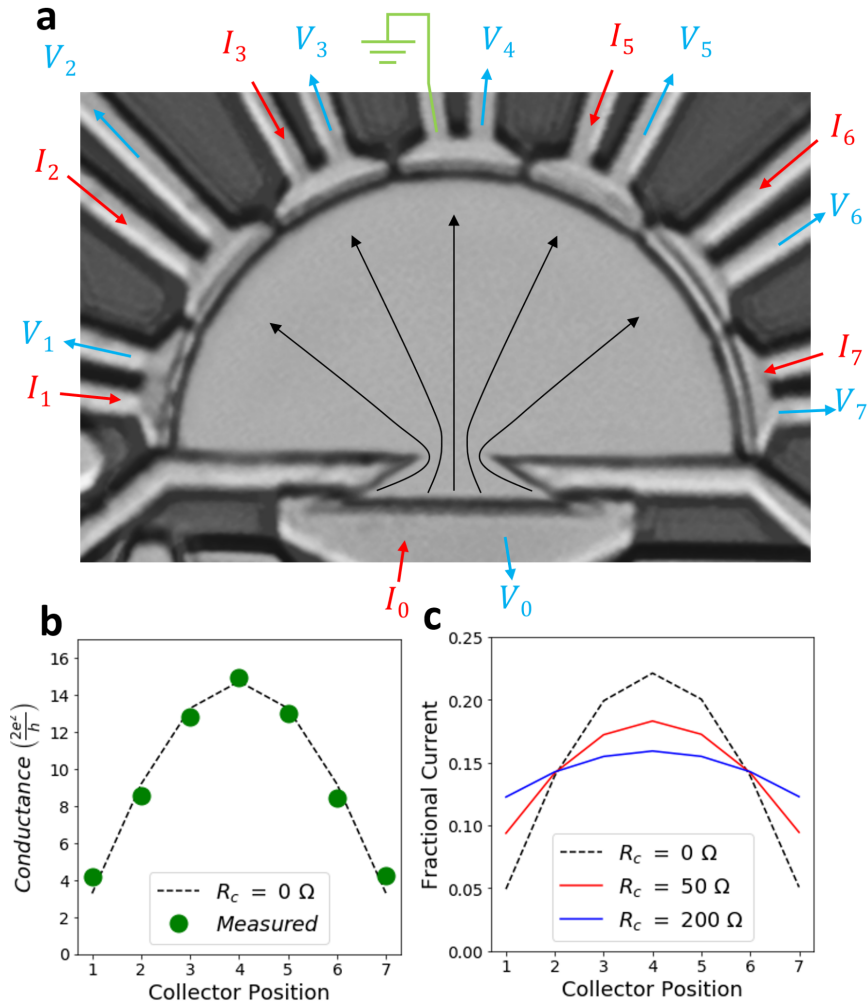

**Figure 4** *Transport device enabled by FBL. (a) A 5 μm radius dome-shaped device is designed to measure the angular flow profile of electrons in hBN-encapsulated graphene at cryogenic temperatures. One source and seven drain contacts each have two electrical leads. For each contact, one lead carries current, and the other serves as a voltage probe. The top contact is physically grounded to prevent the whole device from floating; its current is inferred from the remaining other currents. The drain contacts are virtually grounded by ensuring that $(V_n - V_4) = 0$, effectively eliminating the contact resistance ($R_C$) of the measurement lines and device leads. (b) Measured conductance at 1.4 K as compared to expectations assuming perfect contacts. The dotted line is based on a single parameter fit of the source constriction's resistance. (c) Theoretical flow profiles based on various drain contact resistances. The trace resistances of the device are 30-60 Ohms, and many fridges have filters with resistances ~1 kOhm, making versions of this measurement based not on feedback but on transconductance amplifiers impractical, even if those amplifiers' input offset voltages can be effectively zeroed.*

We tested the device in a liquid helium 1.4 K cryostat. At our measurement temperature, the mean free path of hBN-encapsulated graphene is known to typically exceed the device dimensions ($10 \ \mu m$ diameter semicircle), so transport should be ballistic. In particular, since we are sourcing current through a constriction, we expect to observe behavior consistent with the Sharvin formula--in other words, the expected angular dependence of current density should be cosinusoidal. Measuring at a charge density of $6 \times 10^{11} \ cm^{-2}$ in the graphene, we get a constriction resistance of 196 Ohms, consistent with an effective constriction width of 750 nm. In Fig. 4b, we plot the resulting measured conductances in green circles and the theoretical cosine distribution (appropriately discretized) in black dotted lines. The quantitative agreement is striking.

In a typical transport device measurement, there can be ~10-1,000 Ohm line resistances (depending on whether there are in-line RC filters) and the typical on-chip trace resistance (from wirebond pad to sample region) is ~50 Ohm, meaning that the best possible line resistance is ~50 Ohm with several hundred Ohm being more typical. In Fig. 4c we show the theoretical consequences of finite contact/line resistances. These theoretical results are based on ballistic simulations to determine the conductance matrix of the ideal device and then imposing the specified resistances in series with each contact. With even the best-case scenario, it is evident that the finite line resistances will prevent quantitative measurement of the angular distribution.

The ideal behavior of this device in the ballistic limit demonstrates the technical capability of the feedback lock-in operating in a well-known regime. The more exciting potential for this device and others that rely on virtually-grounded contacts is to probe physics in less-explored regimes such as the onset of electron hydrodynamics and non-linear transport.

## Conclusion
The feedback lock-in is a versatile tool for electrical transport laboratories, both accelerating the pace of discovery by providing parallelizing measurements and enabling new measurement types and geometries. In our implementation, much of the feedback lock-in's capabilities are based on open-source home-built software, granting experimenters the freedom to add capabilities that are otherwise not available/accessible in commercial lock-in amplifiers. Our implementation is scalable, and is ripe for further improvements, with the possibility of greatly enhancing the toolset available to the quantum transport community.

## Data Availability Statement
The data that support the findings of this study are available from the corresponding author upon reasonable request. Detailed component designs will be shared with academic researchers upon request.

## Author Declarations
The authors have no conflicts to disclose.

## Acknowledgements

We would like to thank Sawson Taheri for his advice and help with our preamplifier circuit boards and Aaron Sharpe for his useful feedback. **Funding:** Device fabrication, measurements, and analysis were supported by the U.S. Department of Energy, Office of Science, Basic

Energy Sciences, Materials Sciences and Engineering Division, under contract DE-AC02-76SF00515. Measurement infrastructure was funded in part by the Gordon and Betty Moore Foundation's EPiQS Initiative through grant GBMF3429 and grant GBMF9460. A.W.B. and E.M. were also supported by the Nano and Quantum Science and Engineering Postdoctoral Fellowship at Stanford University. H.S.H. wishes to acknowledge the Mary Gates Endowment scholarship support for this research/project. D.G.-G. gratefully acknowledges support from the Ross M. Brown Family Foundation. Part of this work was performed at the Stanford Nano Shared Facilities (SNSF), supported by the National Science Foundation under award ECCS-1542152.